\documentclass[trackchanges,twocolumn]{aastex701}
\usepackage{CJK}
\usepackage{float}
\usepackage{booktabs}
\usepackage{threeparttable}
\usepackage{amsmath}
\usepackage{xcolor}

\def\eg{{ e.g.,\ }}
\def\ie{{ i.e.,\ }}

\begin{document}
\begin{CJK*}{UTF8}{gbsn}
\correspondingauthor{Long Wang; Zhen Yuan}
\title{The formation of the C-19 progenitor: a primordial cluster heated by gas expulsion}
\author[sname='APJ']{Zhen Wang (王震)}
\affiliation{School of Physics and Astronomy, Sun Yat-sen University, Daxue Road, 519082 Zhuhai, China}
\affiliation{CSST Science Center for the Guangdong-Hong Kong-Macau Greater Bay Area, Zhuhai, 519082, China}
\email{wangzh735@mail2.sysu.edu.cn}  

\author[orcid=0000-0001-8713-0366,sname='APJ']{Long Wang (王龙)} 
\affiliation{School of Physics and Astronomy, Sun Yat-sen University, Daxue Road, 519082 Zhuhai, China}
\affiliation{CSST Science Center for the Guangdong-Hong Kong-Macau Greater Bay Area, Zhuhai, 519082, China}
\email[show]{wanglong8@sysu.edu.cn}

\author[orcid=0000-0002-8129-5415,sname='APJ']{Zhen Yuan (袁珍)}
\affiliation{School of Astronomy and Space Science, Nanjing University, Nanjing, Jiangsu 210093, China}
\affiliation{Key Laboratory of Modern Astronomy and Astrophysics, Nanjing University, Ministry of Education, Nanjing 210093, China}
\email[show]{zhen.yuan@nju.edu.cn}

\author[orcid=0009-0003-7901-5327,sname='APJ']{Jiang Chang (常江)}
\affiliation{Purple Mountain Observatory, CAS, No.10 Yuanhua Road, Qixia District, Nanjing 210034, China}
\email{changjiang@pmo.ac.cn}



\begin{abstract}

The extremely metal-poor nature of the C-19 stream indicates that its progenitor was a primordial stellar system born in the very early Universe. Current observations show that it has a small metallicity dispersion (0.18 at the 95\% confidence level), which is the signature of a globular cluster origin, while at the same time displaying an unusually large velocity dispersion ($\sim10$ km/s) typical of dwarf galaxies. To reconcile this conflicting observational evidence, previous simulations have focused on potential interactions with dark matter subhalos, which can efficiently make a cluster stream dynamically hot. In this work, we explore internal dynamical processes in star cluster formation, focusing on initial conditions shaped by gas expulsion and a top-heavy initial mass function. We find that the large observed velocity dispersion and broad stream morphology can be reproduced by a cluster that underwent severe gas expulsion and expansion during its birth phase, which is potentially a typical formation scenario of extremely metal-poor star clusters. A top-heavy IMF and binaries can also increase the velocity dispersion. The formation of C-19 may involve a combination of these effects.
\end{abstract}


\section{Introduction} 
\label{sec:section 1}

Stellar streams are the relics of globular clusters (GCs) or dwarf galaxies which follow similar orbits after their progenitor systems are tidally disrupted by the galaxy’s potential. In the era of \emph{Gaia} \citep{1-GaiaCollaboration2016, 2-GaiaCollaboration2018, 3-GaiaCollaboration2023}, about 100 stellar streams are discovered and characterized by combining ground-based photometric and spectroscopic observations \citep[see \eg][]{6-Ibata2021, 8-li2022, 9-martin2022a, Ibata2024}. Most of these streams are thin and dynamically cold, originated from dense stellar systems, \ie GCs, compared to wide and hot streams from dwarf galaxies, such as the Sagittarius \citep[see \eg][]{Majewski2003,Belokurov2006,Ibata2020,Ramos2020},
Orphan \citep{Belokurov2007,Koposov2019},
Cetus streams \citep{Newberg2009,Yuan2019,42-yuan2022b}
and LMS-1 \citep{Yuan2020,Malhan2021}.
Streams with very low metallicities, provide precious fossils to study the formation and evolution of their progenitor stellar systems born at very early Universe. Compared to the direct observations from JWST, such as GN-z11, possibly a star cluster born at 440Myr after the big bang \citep[see \eg][]{Bunker2023,Cameron2023},
Galactic Archaeology studies the relics of these primordial systems accreted to the Milky Way. This alternative approach has its own unique advantage: the vicinity of these stellar debris allows us to study their progenitor in a star-by-star manner.

Among all the known low-metallicities streams, the C-19 stream is the most metal-poor stellar structure identified to date \citep{10-martin2022b}, with an average metallicity [Fe/H] $ = -3.38\pm 0.06$ and an extremely small metallicity dispersion ($\sigma_{\rm [Fe/H]} < 0.18$, at 95\% confidence level) via high-resolution spectroscopy. The tiny Fe dispersion as well as the scatter seen in light elements (Na, Mg, Al) strongly indicate its GC origin \citep{13-yuan2022, 12-Bonifacio2024}. However, compared to the other GC streams with similar progenitor mass ($\approx 10^4 ~\rm{M}_{\odot}$), C-19 is usually wide and dynamically hot, which favors a dwarf galaxy origin. The radial velocity dispersion ($\sigma_{\rm v}$) determined from 9 members reaches $6~\mathrm {km~s^{-1}}$ \citep{13-yuan2022}, consistent with results from sub-giants \cite{12-Bonifacio2024}. The tension of its origins is further heightened after doubling the number of confirmed members spanning over 100$^{\circ}$ on the sky \cite{15-yuan2025}. The $\sigma_{\rm v}$ is revised to be $\sim10\,\mathrm {km~s^{-1}}$, whereas the stellar mass remains several 10$^4$ $\rm M_{\odot}$ by counting its bright stars. 

To reconcile the dynamically hot C-19 stream with the assumption of its cluster origin, several studies include the impacts from dark matter. \cite{11-errani2022} explored a scenario that low-density GCs are embedded in a cuspy cold dark matter halo and showed the pre-heating by the halo is able to produce the large $\sigma_{\rm v}$ of C-19. \cite{16-carlberg2024} tested a scenario that streams keep being heated by the subhalos in the MW based on a series of cosmological simulations. This study found that gravitational heating by cold dark matter subhalos over $\sim$ 1.1 Gyr can produce the observed width and $\sigma_{\rm v}$ of C-19, which could explain other hot streams \citep{6-Ibata2021},
such as OmegaCen, M29, NGC288, NGC1851, and NGC1261.

In this work, we try to simulate the C-19 stream on a star-by-star numerical basis. Instead of considering the potential interaction with dark matter halo, we focus on the internal dynamical processes associated with star cluster formation, including  initial mass function, the effects of early gas expulsion, binary fraction, which would critically influence the dynamical state of the stream.

The formation of extremely metal-poor (EMP) star clusters is closely related to their pristine environment. First of all, metallicity may be a factor that determines the initial mass function (IMF) of stars. 
\cite{20-marks2012} studied present-day Galactic GCs and ultra-compact dwarf galaxies and derived their initial conditions, finding systematically top-heavy IMFs in dense and metal-poor environments. The IMF slope becomes flatter with increasing cloud density and decreasing metallicity. \cite{21-chon2021} also showed with hydrodynamic simulations that the IMF varies with metallicity: low metallicity leads to inefficient cooling and favors the formation of massive stars, while higher metallicity supports fragmentation and produces a Chabrier-like IMF \citep{36-chabrier2003}. Secondly, when massive stars form, their feedback can expel gas, suppressing new star formation, reducing gravitational potential, and driving rapid cluster expansion
\citep{Goodwin2006,Bastian2006,18-baumgardt2007,Krause2016,Farias2015,Farias2018,Dinnbier2020a,Pang2020}.
\cite{18-baumgardt2007} studied the effect of gas expulsion with N-body simulations and found that the star formation efficiency (SFE) and the timescale of gas removal determine the survival and expansion of clusters. \cite{Pang2020} found the observed evidence of gas expulsion in NGC 2232 and LP 2439.

The low metallicity of C-19 suggests that it may form with a top-heavy IMF, which would yield many massive stars, whose strong feedback could drive rapid, violent gas expulsion, with their resulting black holes (BHs) providing strong dynamical heating \citep{Giersz2019,Wang2020,Weatherford2021,Rostami2024}. Such feedback can quickly disrupt the cluster, leaving only a dynamically hot stellar stream with no surviving host GC.

To assess how different initial conditions shape the C-19 progenitor cluster and its stream, we run N-body simulations and compare them with observations. The structure of this paper is as follows. In Section~\ref{sec:section 2}, we describe the details of the simulations and the data processing methods used in this work. Section~\ref{sec:section 3} presents the results. Section~\ref{sec:section 4} discusses the results and addresses the limitations of this study as well as future plans. Section~\ref{sec:section 5} provides the conclusions.

\section{Method} \label{sec:section 2} 
\subsection{N-body code}\label{subsec:section2.1}
In this study, we use the high-performance N-body code \verb|PeTar| \citep{22-wanglong2020a,23-wanglong2020b} to simulate the dynamical evolution of star clusters. The code combines the particle-tree particle-particle method \citep{24-Oshino2011} with the slow-down algorithmic regularization (SDAR; \citep{24-wanglong2020c}), which is designed to accurately follow the dynamical evolution of multiple stellar systems. By employing a hybrid parallelization scheme based on the \verb|FDPS| framework \citep{25-iwasawa2016,26-iwasawa2020,27-namekata2018}, the code is capable of handling star-by-star simulations of cluster, allowing us to perform the models required in this work. 

For stellar evolution, we adopt the \verb|SSE/BSE| packages \citep{28-hurley2000,29-hurley2002} and their extended versions \verb|BSEEMP| \citep{30-tanikawa2020}, which provide prescriptions for the evolution of both single and binary stars. In particular, \verb|BSEEMP| offers fitting formulae for the evolutionary tracks of massive stars in extremely metal-poor environments, which satisfy the needs of our simulations of star clusters under such conditions.

In addition, the external tidal field is modeled with the \verb|GALPY| package \citep{31-bovy2015}, where we include the gravitational potential of the Milky Way to account for its influence on the cluster evolution.


\subsection{Galactic potential}\label{sec:potential}

We use the \verb|McMillan17| Milky Way potential \citep{34-mcmillan2017}, comprising bulge, halo, and disk components. 
The Galactic bulge adopts an axisymmetric assumption, with its density distribution based on the parametric model fit by \cite{bissantz2002} to dereddened L-band COBE/DIRBE data \citep{spergel1996}. The dark matter halo density is described with the Navarro--Frenk--White (NFW) profile \citep{Navarro1996}. The Galactic disk is divided into two components: stellar disks and gas disks. The stellar disks' density distribution modeled as
\begin{equation}
\rho_d(R, z) = \frac{\Sigma_0}{2z_d} \exp\left(-\frac{|z|}{z_d} - \frac{R}{R_d}\right),
\label{stellardisk}
\end{equation}
with scale height $z_{d}$, scale length $R_{d}$ and central surface density $\Sigma_{0}$. The parameters for the stellar disks are set using the best-fitting values obtained by \cite{juric2008} from their analysis of Sloan Digital Sky Survey (SDSS: \cite{abazajian2009}) data. The gas disks follow the density law
\begin{equation}
\rho_d(R, z) = \frac{\Sigma_0}{4z_d} \exp\left(-\frac{R_m}{R} - \frac{R}{R_d}\right) \text{sech}^2(z/2z_d),
\end{equation}
which similar to Equation~\ref{stellardisk}, incorporating a central hole with associated scale length $R_{m}$ and an `isothermal' $\text{sech}^2$ vertical profile, with parameters set according to the best-fitting values derived from observational data analyses by \cite{kalberla2008,dame1992,olling2001}.

To obtain the host GC's initial position and velocity in the time-independent Galactic potential, we integrate a centroid test particle from the C-19 stream's present-day phase-space coordinates to approximately 10 Gyr, then reverse its velocity. Forward integration from this initial condition can recover the present-day state. We then use this initial state as the cluster center in the main simulation. The centroid's trajectory  serves as the anchor orbit for the simulated stream and is used to compute stars $\sigma_{\rm v}$. See details of the simulation of the anchor orbit in APPENDIX~\ref{sec:A1}.

\subsection{Models} \label{sec:models}

We explore three different initial conditions of cluster formation in this work: the top heavy IMFs \citep{20-marks2012,Chon2024}, gas expulsion model \citep{18-baumgardt2007}, and 100\% binary fraction \citep{Kropa1995a,Kroupa1995b,Belloni2017}. We will explore these scenarios independently below after setting up a baseline model. In Table~\ref{tab.init}, we summarize all the models with varying parameters. 

\subsubsection{The standard model}
To establish a control group for comparing the heating effects between different models, we initially configured a baseline simulation (the standard model) using the \cite{39-kroupa2001} IMF with masses $0.08\sim150~\mathrm{M_\odot}$ in virial equilibrium, a total mass of \(10^4\) $\mathrm {M_{\odot}}$ {in the same order with the lower mass limit inferred from the updated memberlist \citet{15-yuan2025} and a metallicity of [Fe/H] = $-3.38$ \citep{10-martin2022b}.} 

And we set the binary fraction ($f_{\rm b}$) to 0. We adopt the Plummer model \citep{35-plummer1911} with a typical GC half-mass radius ($r_{\rm h}$) of 1 pc for the initial stellar distribution of clusters. The initial conditions are generated with a modified version of \verb|MCLUSTER| \citep{33-kuepper2011,32-wanglong2019}. 

To further explore the impact of the initial half-mass radius ($r_{\rm h}$) and the initial mass of the cluster mass on the $\sigma_{\rm v}$, we perform two additional simulations based on the standard model: one with $r_{\rm h}$ modified to 3 pc, and another with the initial mass increased to $5\times10^4~\rm{M}_\odot$. The resulting $\sigma_{\rm v}$ at three different epochs are listed in Table~\ref{tab.vary m and rh}. In the standard model, $\sigma_{\rm v}$ at $t=200~\rm Myr$ (after the early rapid stellar evolution around $\sim100~\rm Myr$, when the cluster enters a more steady phase) 
is close to its value at the end of the simulation, indicating that the main change in $\sigma_{\rm v}$ occurs during the early stage of the evolution. All three models remain only partially dissolved at this stage, retaining a bound core. Therefore, the major change in $\sigma_{\rm v}$ occurs between $t=0$ and $t=200~\rm Myr$. Moreover, $\sigma_\text{v}$ at $t = 200$ exhibits a significant decrease compared to its value at $t = 0$. We attribute this to the fact that, within a static Galactic potential and according to Liouville's theorem, $\sigma_\text{v}$ must decrease as the physical width of the stellar stream increases to maintain the conservation of phase-space density \citep[see \eg][]{Helmi1999}.
Comparing the three models, increasing the initial cluster mass leads to a substantial rise in $\sigma_{\rm v}$ at 200~Myr, whereas varying $r_{\rm h}$ has little effect. However, reaching the observed high $\sigma_{\rm v}$ solely by increasing the cluster mass would require the initial mass $\sim10^5 ~\rm{M}_\odot$, well above the observational estimates for C-19 ($3.7\sim5.5\times 10^4~\rm{M}_{\odot}$). Such a massive cluster would not fully dissolve, in conflict with the observations. Due to computational cost, only the standard M1e4 1pc model ($\rm{M} = 10^4~\rm{M}_{\odot}$, $r_{\rm h} = 1$ pc) was evolved to the present day; the two additional models were evolved only to $t=200~\rm Myr$, and we did not explore their long-term evolution.

\begin{table*}[htbp]
\centering
\caption{Dependence of $\sigma_{\text{v}}$ on initial mass and half mass radius in standard model}
\label{tab.vary m and rh}
\begin{tabular}{lccccc}
\toprule
Models & \text{$\sigma_{\text{0,v}}$} ($\mathrm {km~s^{-1}}$) & \text{$\sigma_{\text{200,v}}$} ($\mathrm {km~s^{-1}}$) & \text{$\sigma_{\text{10000,v}}$ } ($\mathrm {km~s^{-1}}$) & \\
\midrule
standard M1e4 1pc & 2.34 & 1.06 & 0.77  & \\
standard M5e4 1pc & 5.16  & 2.58    & \\
standard M1e4 3pc & 1.31 & 0.89   &\\
\bottomrule
\end{tabular}
\tablecomments
{In the model names, M refers to the initial total mass of the cluster in $\rm{M}_{\odot}$, and the value in pc indicates the initial $r_{\rm h}$ of the cluster. $\sigma_{0,\rm v}$, $\sigma_{200,\rm v}$ and $\sigma_{10000,\rm v}$ denote the $\sigma_{\rm v}$ at $t=0$, $t=200 ~\text{Myr}$ and $t=10000~\text{Myr}$ respectively, 
}
\end{table*}

\subsubsection{Top-heavy IMF models}\label{subsubsec:section 2.2.2}

To investigate how varying IMF affects the dynamical evolution of the cluster, we increase the fraction of massive stars by modifying the \cite{39-kroupa2001} IMF, defined as 
\begin{equation}
\xi(m) \propto m^{-\alpha_i} = m^{\gamma_i},
\end{equation}
where
\begin{equation}
\begin{aligned}
\alpha_0 &= +0.3 \pm 0.7, \quad 0.01 \leq m/\rm M_\odot < 0.08, \\
\alpha_1 &= +1.3 \pm 0.5, \quad 0.08 \leq m/\rm M_\odot < 0.50, \\
\alpha_2 &= +2.3 \pm 0.3, \quad 0.50 \leq m/\rm M_\odot < 1.00, \\
\alpha_3 &= +2.3 \pm 0.7, \quad 1.00 \leq m/\rm M_\odot,
\end{aligned}
\end{equation}
and $\xi(m) \, \mathrm{d}m$ is the number of single stars in the mass interval $m$ to $m + \mathrm{d}m$.
 
We consider two cases of modifications on the IMFs:
\begin{itemize}
    \item Include very massive stars by increasing the maximum stellar masses from the standard 150 $\rm M_\odot$ to 500 $\mathrm {M_{\odot}}$ and 1000 $\mathrm {M_{\odot}}$, named as ``$M_{\rm max}500$'' and ``$M_{\rm max}1000$'' models.
    \item Increase the fraction of massive stars by changing the high-mass slope $\alpha_3$ from 2.3 to 1.7 and 2.0, named as ``A1.7'' and ``A2.0'' models.
\end{itemize}

To ensure meaningful comparisons between different models, we adopt the standard model as a benchmark. For the A1.7 model and A2.0 model, the original initial mass would lead to a significantly smaller final stellar number compared to the standard case. Therefore, we rescale their initial total stellar number so that the final stellar number is comparable to that of the standard model. This adjustment results in different initial mass of the cluster, as listed in Table~\ref{tab.varyimf}.

A top-heavy IMF produces more numerous and more massive BHs from massive-star evolution. Table~\ref{tab.varyimf} reports the maximum and total BH masses after 30 Myr of evolution under standalone stellar evolution using \verb|BSEEMP|. As the table shows, the ratio of $M_{\text{30,BHtot}}$ to $M_{\text{30,tot}}$ ($q$) in $M_{\rm max}500$ and $M_{\rm max}1000$ models doesn't change a lot compare with the standard model, but $M_{\text{30,BHmax}}$ is highest in the $M_{\rm max}1000$ model. The A1.7 and A2.0 models show a significant improvement in the $q$ value compared to the standard model. Except for the $M_{\rm max}500$ model, the mass and fraction of BHs in the other models are all enhanced compared to the standard model. Therefore, BH heating should be stronger for top-heavy IMFs, accelerating cluster disruption \citep{Wang2020}, and potentially altering stream morphology and $\sigma_{\rm{v}}$.

\begin{table*}[htbp]
\centering
\caption{Maximum and total BH and all object masses for Top-heavy IMF models.
}
\label{tab.varyimf}
\begin{tabular}{lccccc}
\toprule
Models & \text{$M_{\text{0,tot}}$} ($\mathrm {M_{\odot}}$) & \text{$M_{\text{30,tot}}$} ($\mathrm {M_{\odot}}$) & \text{$M_{\text{30,BHtot}}$} ($\mathrm {M_{\odot}}$) & \text{$M_{\text{30,BHmax}}$} ($\mathrm {M_{\odot}}$) & \text{$q$} \\
\midrule
standard & 10000 & 8497.6 & 693.9 & 40.5 & 0.082 \\
$M_{\text{max}}$500 & 10000 & 8626.7 & 431.3 & 32.5 & 0.049 \\
$M_{\text{max}}$1000 & 10000 & 8541.7 & 735.5 & 286.4 & 0.086 \\
A2.0 & 20153 & 14254.9 & 2400.1 & 84.1 & 0.168 \\
A1.7 & 40900 & 23373.0 & 7656.0 & 45.7 & 0.328 \\
\bottomrule
\end{tabular}
\tablecomments
{$M_{\text{0,tot}}$ is the initial total mass of the cluster, and the $M_{\text{30,tot}}$ is the total mass of the cluster at 30 Myr, $M_{\text{30,BHtot}}$ and $M_{\text{30,BHmax}}$ mean the total mass and the maximum mass of BHs in the cluster at 30 Myr, and the $q$ is the ratio of $M_{\text{30,BHtot}}$ to $M_{\text{30,tot}}$.}
\end{table*}

\subsubsection{Gas expulsion models}\label{subsubsec:section 2.2.3}

Radiation pressure and stellar winds from massive stars expel the primordial gas, suppress further star formation, and weaken the system's gravitational potential, strongly influence the final properties and survival probability of clusters. This gas expulsion phase is brief, lasting less than a dynamical time. The strength of this process depends on the SFE, which is the fraction of gas converted into stars and defined as follows: 
\begin{equation}
 \epsilon=\frac{M_{\star}}{M_{\star}+M_{\rm g}},
  \label{Eq.SFE}
\end{equation} 
where $M_\star$ and $M_{\rm g}$ are the total stellar and gas masses before gas expulsion, and after all stars have formed. If the SFE falls below a critical threshold, the cluster becomes supervirial, leading to significant stellar mass loss and potentially complete cluster dissolution. According to \cite{18-baumgardt2007}, under instantaneous gas removal, clusters require $\epsilon>33\%$ to survive gas expulsion. This survival threshold decreases substantially for longer gas removal timescales. In the adiabatic limit and under weak external tidal fields, clusters can survive with $\epsilon$ as low as $10\%$. Surviving clusters typically undergo significant expansion due to gas expulsion. 
C-19 has extremely low metallicity. \cite{Chon2024} shows that for [Z/H] in between $-3$ and $-4$, the SFE is below 0.1, implying that C-19 may undergo significant gas expulsion and become unbound soon after star formation.

Our simulations do not include gas dynamics. To model rapid dissolution via gas expulsion, we assume instantaneous gas loss: the stars' kinetic energy remains unchanged, but the removal of gas reduce the gravitational potential, driving the cluster out of virial equilibrium into a supervirial state. We represent this by scaling the initial stellar velocities in the standard model by a factor $s_{\rm v}>1$. 

If the system is virialized before gas expulsion, the stellar potential satisfies
\begin{equation}
    \Phi_0 = s_{\rm v}^2 \Phi_1,
    \label{eq:phi0to1}
\end{equation}
where $\Phi_1$ denotes the post-expulsion potential.
This scaling provides a rough estimate of the SFE. 
Assuming gas and stars share the same density profile, $\Phi_0 \propto M_\star + M_{\rm g}$, and after $\Phi_1 \propto M_\star$, yielding
\begin{equation}
    \epsilon = \frac{\Phi_1}{\Phi_0} = \frac{1}{s_{\rm v}^2}
\end{equation}

We consider $s_{\rm v}=$ 2.0, 3.0, 4.0 and 5.0 based on the standard model, labeled ``V2.0",``V3.0", ``V4.0" and ``V5.0" in the model names, respectively. 
Since increasing the initial cluster mass raises $\sigma_{\rm v}$ (Table~\ref{tab.vary m and rh}), we also perform three higher-mass gas-expulsion models: M5e4 ($5\times10^4 M_\odot$) with V3.0 and V4.0, and M2e4 ($2\times10^4 M_\odot$) with V4.0, to investigate their combined effects.

To further indicate how the modeling initial conditions lead to different stream kinematics, we present the time evolution of the virial ratio for all models in APPENDIX~\ref{sec:A2}.
\subsubsection{The binary model}\label{subsubsec:section 2.2.4}

When observing and confirming member stars and measuring their radial velocity ($v_{\rm r}$), binary systems can also cause measurement errors in $v_{\rm r}$ due to their impact on spectral broadening. In addition, dynamical interactions between binaries and other stars may also increase $\sigma_{\rm v}$. To investigate these effects, we modify the standard model by introducing primordial binaries with $f_{\rm b}=100\%$ (named as ``B1.0").  

For massive binaries, the orbital parameters follow the observational OB binary distributions from \cite{Sana2012}. For low-mass binaries, we adopt the Kroupa primordial binary model \citep{Kropa1995a,Kroupa1995b,Belloni2017}, which assumes a universal period distribution spanning wide separations, random mass-ratio pairing, and a thermal eccentricity distribution. Recent studies estimate GC binary fractions of $f_{\rm b}\sim 6\% - 15\%$ \citep{Bellazzini2002,Sollima2007,Milone2008,Sommariva2009,Lucatello2015,Dalessandro2018}. Although our initial $f_{\rm b}=100\%$ is extreme, most wide binaries are dynamically unstable and are rapidly disrupted, leaving a much lower final fraction. This setup therefore provides an upper limit on how binaries may bias $v_{\rm r}$ measurements and inflate $\sigma_{\rm v}$.

\begin{table*}[htbp]
  \centering
    \caption{Initial conditions and parameter variations of all simulation models}
    \label{tab.init}
    \begin{tabular}{lcccccccc}
      \toprule
      Models & $M_{0,\text{tot}}$ ($\mathrm{M_{\odot}}$) & $M_{\text{max}}$ ($\mathrm{M_{\odot}}$) & $\alpha_3$ & $s_{\rm v}$ & $f_{\rm b}$ & $N_{\text{tot}}$ & $N_{\text{sel}}$ & \text{$t_{\text{dis}}$} (\text{Myr}) \\
      \midrule
      standard    & 10000 & 150  & 2.3 & 1.0 & 0 & 16701 & 494 & not  \\
      $M_{\text{max}}$500  & 10000 & 500  & 2.3 & 1.0 & 0 & 16190 & 456 & not \\
      $M_{\text{max}}$1000 & 10000 & 1000 & 2.3 & 1.0 & 0 & 16700 & 526 & not \\
      A2.0    & 20153 & 150  & 2.0 & 1.0 & 0 & 20000 & 693 & 2600 \\
      A1.7    & 40900 & 150  & 1.7 & 1.0 & 0 & 20000 & 755 & 2452 \\
      V2.0    & 10000 & 150  & 2.3 & 2.0 & 0 & 16345 & 612 & 12 \\
      V3.0    & 10000 & 150  & 2.3 & 3.0 & 0 & 17392 & 536 & 8 \\
      V4.0    & 10000 & 150  & 2.3 & 4.0 & 0 & 15986 & 370 & 4 \\
      V5.0    & 10000 & 150  & 2.3 & 5.0 & 0 & 17033 & 296 & 7 \\
      V3.0 M5e4    & 50000 & 150  & 2.3 & 3.0 & 0 & 86811 & 1271 & 5  \\
      V4.0 M2e4    & 20000 & 150  & 2.3 & 4.0 & 0 & 33499 & 517 & 4 \\
      V4.0 M5e4   & 50000 & 150  & 2.3 & 4.0 & 0 & 84828 & 848 & 5 \\
      B1.0  & 10000 & 150  & 2.3 & 1.0 & 1 & 16974 & 402 & not \\
      \bottomrule
    \end{tabular}
    \tablecomments 
    {$M_{\text{0,tot}}$ is the initial total mass of the cluster, $M_{\text{max}}$ is the maximum stellar mass of IMF, $\alpha_3$ is the high-mass slop of the IMF, $s_{\rm v}$ is  the velocity expansion factor, the $f_{\rm b}$ means the the binary fraction is different clusters. 
    $N_{\rm tot}$ is the total stellar number, $N_{\rm sel}$ is the selected stellar number after data filtering in different clusters, \text{$t_{\text{dis}}$} represents the dissolution times of the cluster cores.}
\end{table*}

\subsection{Preparations for the comparisons with observations} \label{subsec:section 2.3}

For a proper simulation-observation comparison, we restrict the observational field, remove undetectable stars (e.g., faint stars and compact objects), and account for small-number statistics in the simulation data using the following three filtering steps.

Based on the spatial distribution of C-19 stream members reported by \cite{15-yuan2025}, we first apply a spatial filter to the simulated data in the RA-Dec plane, selecting stars within $\mathrm{RA}\in[212^\circ,360^\circ]$ and $\mathrm{Dec}\in[-65^\circ,77^\circ]$. To mitigate projection effects in which multiple wraps overlap in the 2D RA-Dec projection, we further refine the selection using the anchor orbit by requiring the 3D velocity vectors of simulated stars to be within $30^\circ$ of the orbital velocity direction and by selecting stars located within a fixed 3D distance from the orbit, thereby isolating a stream wrap most representative of the observations. For the B1.0 model, an additional filtering step is applied: in this simulation, the cluster core survives and produces an artificially high $\sigma_{\rm v}$ near the core region; since no such undisrupted core is observed, these stars are manually excluded when computing $\sigma_{\rm v}$ to ensure a consistent comparison. We then convert stellar properties into Gaia-mock G-band magnitudes using \verb|GALEVNB| \citep{40-pang2016} and apply an observational magnitude cut of $G\in[13.4,17.7]$. Finally, to account for small-number statistics in the observations, the simulated stream is divided into three spatial regions:
\begin{itemize}
    \item Region A: $\mathrm{RA}\in[352^\circ,357^\circ]$ and $\mathrm{Dec}\in[20^\circ,32^\circ]$ with 10 stars; 
    \item Region B: $\mathrm{RA}\in[350^\circ,355^\circ]$ and $\mathrm{Dec}\in[-2^\circ,9^\circ]$ with 4 stars; 
    \item Region C: $\mathrm{RA}\in[346^\circ,348^\circ]$ and $\mathrm{Dec}\in[-16^\circ,-9^\circ]$ with 3 stars,
\end{itemize}
In each region, the same number of stars as observed are randomly selected; this procedure is repeated ten times, and the resulting $\sigma_{\rm v}$ values are averaged to reduce statistical fluctuations. For Region C, due to the limited number of samples within the original observational range, we expanded the $\mathrm{Dec}$ range by extending the lower bound from $-11^\circ$ to $-16^\circ$ to improve the statistical robustness.

For the V3.0 M5e4, V4.0 M2e4, and V4.0 M5e4 models, the increased initial mass results in a spatial extent significantly larger than that covered by the observational data. To ensure a consistent width measurement, we expand the three spatial regions perpendicular to the stream in all models to include all nearby simulated stars.



After applying the above filtering and mock-observation procedures uniformly to all simulations, we examine the number of stars retained in the observed region, $N_{\rm sel}$, as a consistency check to ensure that differences in the measured properties across models are not primarily driven by large difference in star's number.

Taking the standard model as a benchmark, we require 
$N_{\rm sel}$ to remain at a comparable level across models that share the same initial total mass. This excludes the three gas expulsion models with modified initial mass of the cluster (V3.0 M5e4, V4.0 M2e4, V4.0 M5e4; listed in Table~\ref{tab.init}), which are designed to explore the impact of higher initial mass on $\sigma_{\rm v}$ rather than to maintain comparable $N_{\rm sel}$.
For the remaining cases, the binary model generally remains consistent with the standard model, the $N_{\rm sel}$ of the two top-heavy models both increase with the increase of massive stars, while the $N_{\rm sel}$ for the gas expulsion models show considerable fluctuation and exhibit a decreasing trend as $s_{\rm v}$ increases. This decline occurs because higher $s_{\rm v}$ values cause the gas expulsion models to become more spatially diffuse, leaving fewer stars within the sampling region by the final stage of evolution.

\section{RESULT} \label{sec:section 3} 


\subsection{The velocity dispersion} \label{subsec:section 3.1}

\begin{figure*}
    \centering
    \includegraphics[width=1\linewidth]{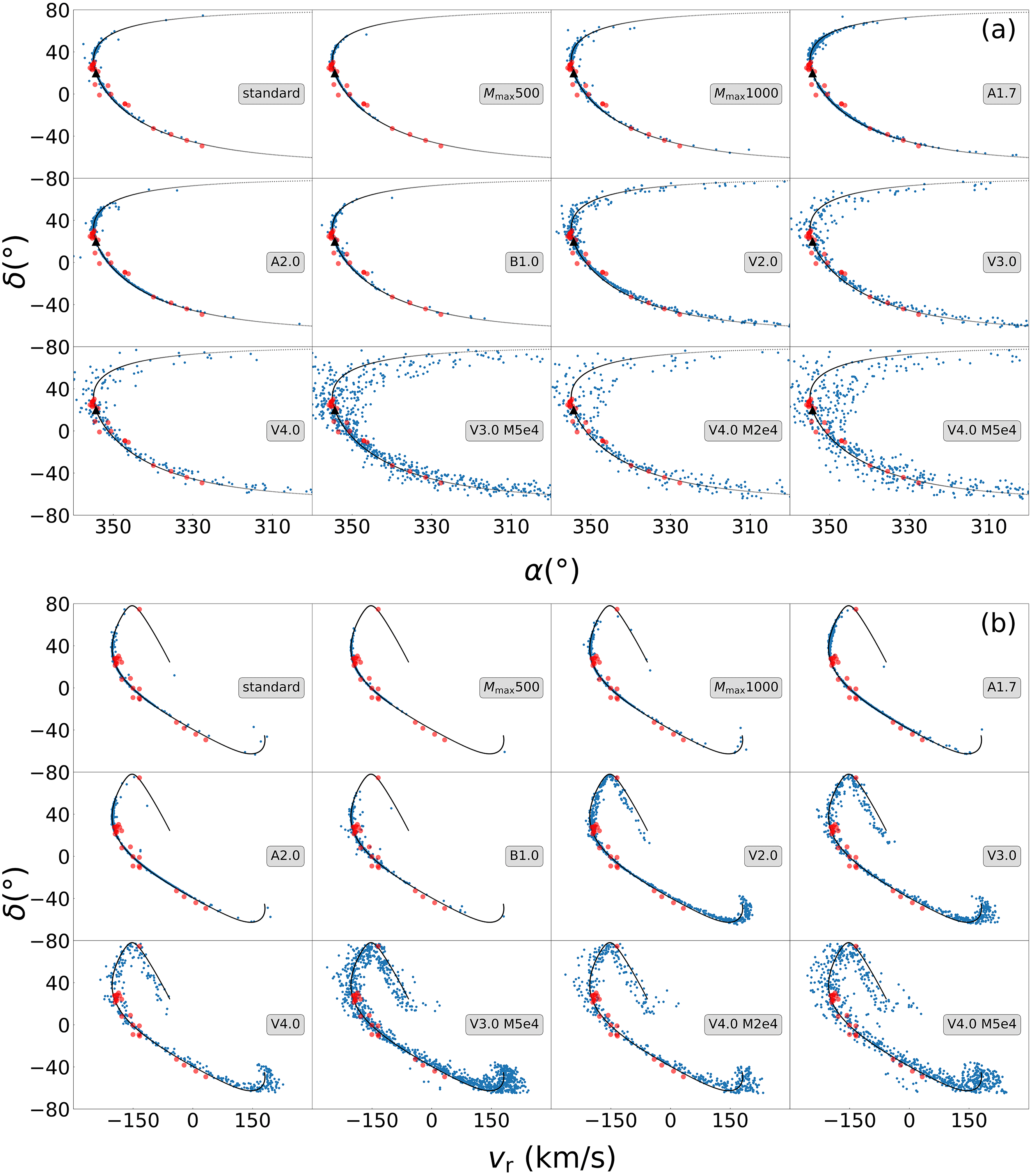}
    \caption{(a) Spatial distribution of stars in the C-19 stream. (b) $v_{\rm r}$ versus $\delta$ for C-19 stars. Each panel shows a different simulation model. Grey dots represent the full stream, blue dots indicate the selected stellar used for comparison with observation, red dots mark the observed stars, black dots denote the orbit, and black triangles represent the present-day position of the stream's centroid.
 }
    \label{fig.2}
\end{figure*}

In the on-sky projection in Fig.~\ref{fig.2}, for each simulated star, we locate the nearest point on the trajectory (by Euclidean distance) and compute the star's velocity relative to that point. We then calculate the dispersion of the selected stars in each stream region. 


Using our centroid orbit as the reference frame, we recalculate the $\sigma_{\rm v}$ and obtain 12.6 $\mathrm {km~s^{-1}}$, using 17 members from \cite{15-yuan2025} by ignoring the northern and southern member located on the opposite of the disk. This recalculate dispersion serves as the reference value against which $\sigma_{\rm v}$ from all subsequent models are evaluated.

Figure~\ref{fig.2} presents the distribution of C-19 stream stars in $\alpha$-$\delta$ and $v_{\rm r}$-$\delta$ spaces, comparing observational and simulated data across different models.  Compared with the standard model, the gas expulsion models produce a more extended spatial morphology and stronger dynamical heating, both of which increase with larger $s_{\rm v}$. Increasing the initial cluster mass further amplifies these effects in both morphology and $\sigma_{\rm v}$. These simulations reproduce the full spatial extent and density distribution of observed member stars. 

The simulated stream members  shown in Figure~\ref{fig.2} only cover a single wrap that overlaps with the observations, whereas the full stream extends much farther. As a result, the selected streams produced by different models exhibit similar morphologies. For gas-expulsion models, although the cluster dissolves within $\sim10~\text{Myr}$ due to gas expulsion, its central region expands gradually rather than dispersing instantaneously. Therefore, even after $10~\text{Gyr}$, a relatively dense group of stars with coherent motion can still remain near the original cluster center, instead of leaving an empty region as would be expected in the case of instantaneous disruption. 

The V4.0 model was adopted by \cite{venn2026} to support a primordial origin for the C-19 stellar stream. This simulated stream provides dynamical evidence that a low-SFE event can indeed evolve into the observed stream following $\sim$10 Gyr of tidal interaction with the Milky Way.

In contrast, the top-heavy IMF models exhibit narrower, dynamically colder streams, similar to the standard case. Among them, the $M_{\max}1000$ model shows a slightly stronger expansion than the $M_{\max}500$ case, but both remain less extended than the $\alpha_3$-modified models (A1.7 and A2.0). All top-heavy models are only marginally more diffuse than the standard model. 

The binary model (B1.0) doesn't show a puffier morphology but it shows a higher $\sigma_{\rm v}$ than the standard model and exhibits a pretty high peak nearby $\delta = 20^\circ$ where is its undisrupted core in $v_{\rm r}$-$\delta$ space. Previous studies have suggested that binary-mediated interactions can preferentially eject more massive stars at higher velocities, potentially leading to a higher observed velocity dispersion for brighter stream members \citep{Grondin2024,Weatherford2026}. To examine this in our models, we lowered the maximum Gaia G-band magnitude for membership selection from $G_{\rm max}=17.7$ to $G_{\rm max}=17.0$, thereby selecting a brighter subset of stars. We find that the resulting $\sigma_{\rm v}$ decreases from 0.86 to 0.74 $\mathrm{km~s^{-1}}$, indicating that the enhanced $\sigma_{\rm v}$ in B1.0 models is not primarily driven by massive, binary-ejected stars, but instead is primarily driven by the internal orbital motions of binary components.

Figure~\ref{fig.3} (a) presents the $\sigma_{\rm v}$ from all simulations. The gas expulsion models demonstrate a pronounced large $\sigma_{\rm v}$. Even at the lowest expand factor as 2.0, the $\sigma_{\rm v}$ exceeds that of the standard simulation. The magnitude of the $\sigma_{\rm v}$ increases markedly with higher expand factors. Crucially, a part of simulations employing an expand factor of 3.0, 4.0 and 5.0 yield a $\sigma_{\rm v}$ that closely matches the observational result. Compared with their corresponding baseline models, all higher-mass M2e4 and M5e4 models exhibit a more pronounced increase in $\sigma_{\rm v}$ and can match the observational result of 12.6 $\mathrm {km~s^{-1}}$.

In addition, the estimated SFE for model V3.0-V5.0 lies in the range from $1.1\times10^{-1}\sim4\times10^{-2}$, consistent with the low-metallicity SFE from \cite{Chon2024}.




Regarding our explorations of variations of IMFs and binary fractions, 
$\sigma_{\rm v}$ calculated for the two kinds of top-heavy IMF models doesn't exceed the standard model, but both show a trend of increase. Similarly, the binary B1.0 model produces stronger dynamical heating than both the standard and top-heavy IMF models, but its $\sigma_{\rm v}$ remains below the observed value. These results suggest that neither top-heavy IMFs nor binary heating can fully account for the observed dynamical state of the C-19 stream. 






\subsection{Width} \label{subsec:section 3.2}
Using the anchored orbit as a reference, we fit the stream distributions of both observational and simulated data in the direction perpendicular to the orbit with a Gaussian function, and calculated the corresponding stream widths for comparison. 

\begin{figure*}
    \centering
    \includegraphics[width=1\linewidth]{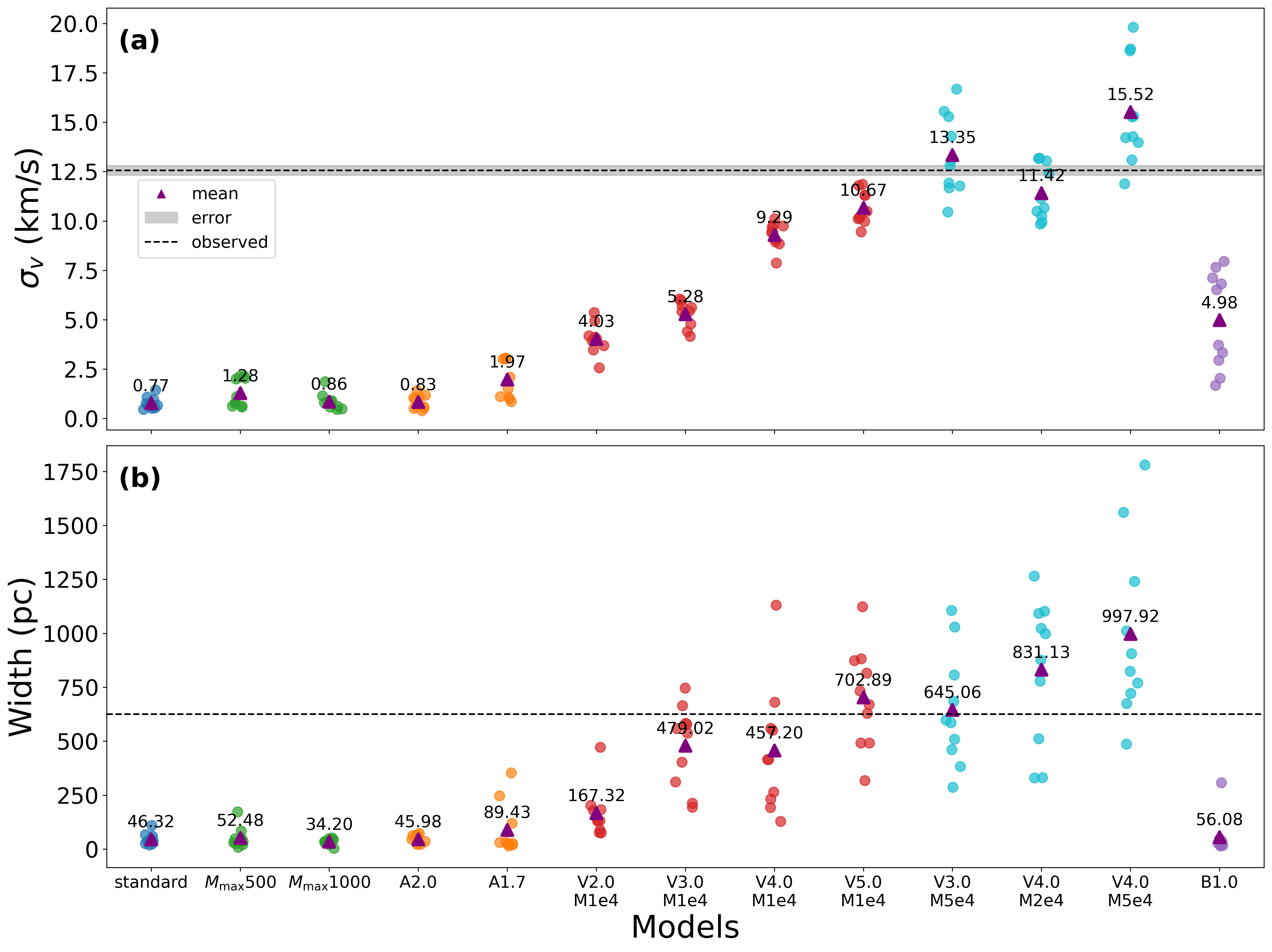}
    \caption{(a) The measured $\sigma_{\rm v}$ values for each model using ten subsamples. (b) The measured width values for each model using ten subsamples. Colors indicate different groups of models. Purple triangles indicates the average $\sigma_{\rm v}$ and width of the subsamples. The black dashed line represents the observed $\sigma_{\rm v}$  and width reference, the grey region represents the error of the observed $\sigma_{\rm v}$.
    }
    \label{fig.3}
\end{figure*}

Figure~\ref{fig.3} (b) presents the widths for all models. The widths of the gas expulsion models increase with $s_{\rm v}$, and higher initial cluster masses also lead to noticeably wider streams. Among these models, M1e4 with V5.0 and M5e4 with V3.0 matches the observed stream width of 625.74 pc, and the higher-mass M2e4 and M5e4 models with V4.0 significantly exceed this value.

In contrast, most of the top-heavy IMF models yield widths similar to the standard model, except for the $\alpha_3$-modified model A1.7, which yields a larger value. The widths of the binary model behave differently from its $\sigma_{\rm v}$, with a marginal increase relative to the standard model. 

Overall, the width trends broadly track $\sigma_{\rm v}$ in Figure~\ref{fig.3} (a), indicating that models with larger $\sigma_{\rm v}$ also produce more spatially extended streams. However, the initial cluster mass has a stronger effect on width than $\sigma_{\rm v}$, likely because $\sigma_{\rm v}$ is measured in observationally constrained regions and thus has an upper limit, as higher-velocity stars preferentially move outside these regions. 

Although M2e4 and M5e4 models predicts streams wider than observed, they are not fully excluded, as C-19 stream may extend beyond current samples because  distant members are difficult to distinguish from background contamination.


\subsection{The best-match model} \label{subsec:section 3.3}

Considering both $\sigma_{\rm v}$ and stream width, the best agreement with observations is achieved in the gas expulsion scenario with M1e4-V5.0 and M5e4-V3.0 models. In this regime, the initial stellar velocities are amplified to three to five times the virial equilibrium value, corresponding to a SFE of approximately $1.1\times10^{-1}\sim4\times10^{-2}$. This leads to rapid heating, expansion, and disruption of the cluster, ultimately producing a wide stellar stream whose $\sigma_{\rm v}$ and width are both consistent with the observed values. Variations in the $r_{\rm h}$ have negligible impact, and the top-heavy IMF and binary models are unable to reproduce the observed kinematics or spatial extent.

Our simulations further show that the stream becomes dynamically hot at very early times, around 1 Gyr, and by $\sim 3$ Gyr it already exhibits a length and width comparable to those produced after 10 Gyr of evolution. This indicates that the observed kinematic state of the stream does not require a long disruption timescale and can be achieved within only a few Gyrs.


\section{DISCUSSION} \label{sec:section 4} 

We compare different proposed mechanisms for the origin of the high $\sigma_{\rm v}$ and clarify the advantages and limitations of our model. In the preheating framework of \cite{11-errani2022}, the high $\sigma_{\rm v}$ is reproduced by assuming a dark matter–dominated progenitor, but this relies on a non-standard scenario for GCs. In contrast, the external-heating model of \cite{16-carlberg2024} invokes perturbations from dark matter halos and Galactic substructure; however, such interactions would likely also heat many other GC streams, which are generally observed to remain dynamically cold. In addition, we note that the GC models adopted in \cite{11-errani2022} and \cite{16-carlberg2024} tend to assume somewhat larger characteristic sizes than those typically inferred for Galactic GCs, with half-light or half-mass radii of 20 pc and 4 pc, respectively. 
In our model, the high $\sigma_{\rm v}$ is primarily attributed to early gas expulsion, while we adopt a more compact and observationally motivated initial $r_\text{h}$ of 1 pc. Furthermore, the top-heavy IMF and high binary fraction could also contribute to the heating of the stream.
These mechanisms provide sources of intrinsic, global heating without requiring dark matter or strong external perturbations, and the gas expulsion scenario in particular is consistent with the observed stream morphology. However, these interpretations are based on simulations performed in a simplified, static Galactic potential; therefore, it remains essential to test their robustness in more realistic, time-evolving environments.

In our study of the possible formation mechanism of the C-19 stream, we use an approximate static Galactic potential model and do not include the long-term evolution of the Milky Way. Although Galactic evolution could non-negligibly affect the stream dynamics, especially $\sigma_{\rm v}$, our results show that initial gas expulsion alone is significant and can reproduce the observed $\sigma_{\rm v}$ and width of the C-19 stream. Thus, while a time-evolving potential is not explored in detail, this simplification does not diminish the importance of gas expulsion. 

In addition, phase mixing in a realistic, evolving, and substructured Galactic potential may restrict identifiable stream members to relatively recent ejecta \citep[last few Gyr; e.g.][]{Johnston2002,Panithanpaisal2026,Carlberg2009,kupper2015,Pearson2017,Bonaca2019,Woudenberg2023}. In our models, gas-expulsion driven dissolution occurs within $\sim10$~Myr, so the resulting stream may already be largely undetectable, though some fraction could remain observable. Although disrupted by gas expulsion, the cluster's central region expands gradually rather than dispersing instantaneously. Consequently, even after 10~Gyr, a relatively dense, co-moving group persists near the original center, rather than leaving an empty region, as seen in the simulated stream members overlapping with observations in Figure~\ref{fig.2}. Tracing these members back 1.5 Gyr shows they were more concentrated, though already elongated. While phase mixing likely renders most of the stream undetectable, a dense component near the original cluster center may persist, suggesting that part of the central stream could remain observable.

We also explored stream evolution with and without gas expulsion in a time-dependent Milky Way-like potential from \citet{Ishchenko2023}, based on the TNG100 run of the IllustrisTNG cosmological simulation. These preliminary results are not included here. Compared to a static potential, the time-dependent case produces a more extended stream after 10 Gyr, though it remains partially detectable. Future work will examine how time-dependent potentials affect stream velocity dispersion under different initial conditions.

The process of gas expulsion is closely related to the SFE, and the value of $\epsilon$ directly influences the evolution of a cluster after gas expulsion. Our setting of $\epsilon$ is simplified. 
In future work, more detailed hydrodynamical simulations could be used to model the process of gas expulsion and its interaction with star formation in more detail, to obtain a more reliable estimate of $\epsilon$. This would help to evaluate quantitatively how gas expulsion under different SFE conditions affect the cluster structure and $\sigma_{\rm v}$.

In the B1.0 model, the surviving binary fraction of stripped stars lies in the range of $\sim47\%$ to $\sim88\%$, with a dominance of intermediate period systems ($\sim$ 1 -- 1000 years). 
Future observations capable of identifying binary candidates in the C-19 stream will be essential for constraining its primordial binary fraction and testing these predictions.


\section{CONCLUSION} \label{sec:section 5} 
In this study, we use the N-body code \verb|PeTar| to study the origin of the C-19 stream, assuming it is originated from a GC. We explore the influence of top-heavy IMFs, initial gas expulsion conditions as well as high binary fractions in GC. Our simulations show that in the scenario of early gas expulsion, producing a supervirial initial state with a virial ratio equal 25 for a $1\times10^4~\rm{M}\odot$ cluster or equal 9 for a $5\times10^4~\rm{M}\odot$ cluster, rapidly heats and expands the system, yielding a wide stream and a $\sigma_{\rm v}$ consistent with the observed value of $\sim 12.6~\mathrm {km~s^{-1}}$. Top-heavy IMFs have little effect on the stream's morphology or dynamics. Binary system can lead to a higher $\sigma_{\rm v}$ than the standard model and has little effect on the stream's morphology; however, the resulting stream is too thin compared to the observations.

Overall, the formation of the C-19 stream is likely to be a complex process. While our results highlight gas expulsion as an efficient mechanism for producing the required dynamical heating, it is not necessarily the only viable explanation. A combination of effects, such as the interplay between a top-heavy IMF, a high binary fraction and a time-dependent Galactic potential, could also in principle contribute to enhanced heating under certain conditions.

These results suggest that gas expulsion alone is sufficient for the dynamical heating of C-19, without requiring a dwarf galaxy origin or a GC heated by dark matter halos. The derived SFE, $\epsilon \le 6.25\times10^{-2} $ (from model V4.0), is consistent with the predicted range for SFE in low-metallicity environments \citep{Chon2024}. 

This finding gives insight into the rarity of extremely metal-poor GCs: gas expulsion may occur more easily in low-metallicity environments, heating and dissolving dense star clusters early on to form diffuse streams instead of long-lived dense clusters. Consequently, while  extremely metal-poor dense GCs are rare in the present Milky Way, more stellar streams likely exist that formed under similar conditions, which are expected to be discovered with the upcoming \emph{Gaia}DR4, WEAVE \citep{Jin2024}, and 4MOST \citep{de2019}. Combining tailored simulations that take into account the internal dynamical processes would help us constrain the physical processes in the formation of primordial star clusters as well as their dynamical evolution in the Milky Way.

\section{Software and third party data repository citations} \label{sec:cite}
\software{numpy (\citealp{85-Harris.2020}),
          matplotlib (\citealp{86-Hunter.2007}),
          \texttt{sdar} (\citealp{24-wanglong2020c}, https://github.com/lwang-astro/SDAR),
          \texttt{petar} (\citealp{22-wanglong2020a}, https://github.com/lwang-astro/PeTar),
          \texttt{mcluster} (\citealp{33-kuepper2011}, https://github.com/lwang-astro/mcluster)
          \texttt{Galpy} (\citealp{31-bovy2015}, https://github.com/jobovy/galpy)
          }

\begin{acknowledgments}
Z.W., Z.Y. and L.W. appreciate the insightful discussions about the C-19 stream with Nicolas F. Martin and Kim Venn.
L.W. thanks the support from the National Natural Science Foundation of China through grant 12573041 and 12233013, the High-level Youth Talent Project (Provincial Financial Allocation) through the grant 2023HYSPT0706,  the Fundamental Research Funds for the Central Universities, Sun Yat-sen University (2025QNPY04). Z.Y. is supported by the National Key R\&D Program of China via 2024YFA1611601. J.C. acknowledges the support of the National Natural Science Foundation of China No. 12573024 and 12533007. The authors acknowledge the Beijing Beilong Super Cloud Computing Co., Ltd for providing HPC resources that have contributed to the research results reported within this paper.URL: http://www.blsc.cn/. 

 \end{acknowledgments}

 \begin{contribution}
Z.W. was responsible for performing the simulations, data analysis, and writing the manuscript.
L.W. developed the initial research concept, contributed to discussions, edited the manuscript, and supervised the overall project.
Z.Y. developed the observational-comparison concept, provided observational data, contributed to discussions, edited the manuscript, and co-supervised the project.
J.C. contributed to the discussions and edited the manuscript.




\end{contribution}

\appendix

\section{Orbit simulation} \label{sec:A1}
\renewcommand{\thefigure}
{A\arabic{figure}} 
\setcounter{figure}{0}

\begin{figure}[ht!]
    \centering
    \includegraphics[width=1\linewidth]{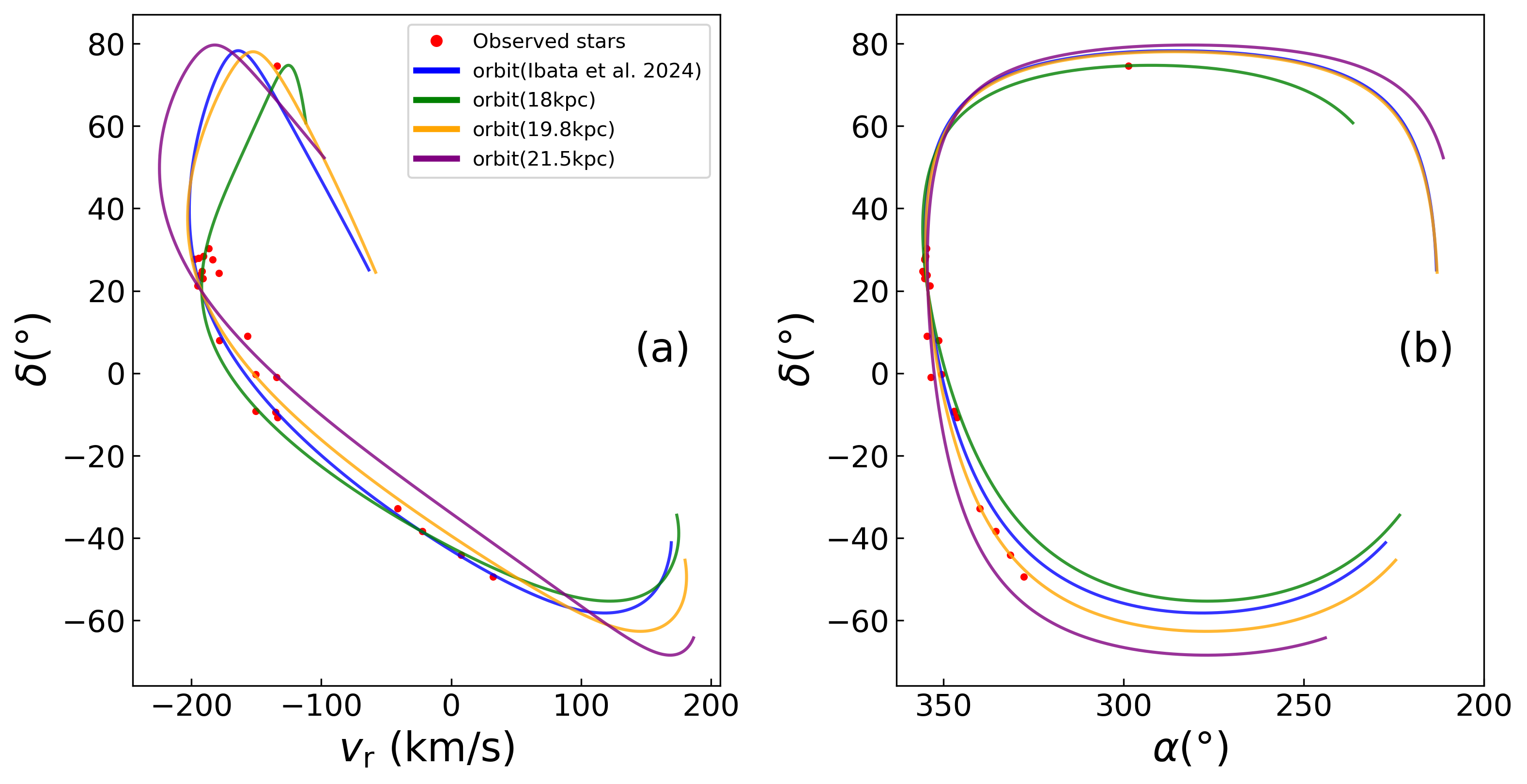}
    \caption{
    The simulation of orbit. (a) The Orbit in the ($v_{\rm r}$,$\delta$) space. The red dots mark the observed stars. The blue line represents Ibata's orbit, the green, black, and purple lines represent the orbits of the stream at distances of 18 kpc, 19.8 kpc and 21.5 kpc, respectively. (b) Spatial distribution of the orbit. As shown in the figure, the orbits exhibit different bending trends in $\alpha$-$\delta$ and ($v_{\rm r}$,$\delta$) spaces as the distance increases. Considering these factors, we adopt the orbit at a distance of 19.8 kpc as the anchor orbit.}
    \label{A1}
\end{figure}

Using test-particle simulations of the cluster's centroid as described in Section~\ref{sec:potential}, we derive the orbit presented in Figure~\ref{A1}. While \cite{10-martin2022b} estimated the C-19 center at 18 kpc and  \cite{14-Viswanathan2024} corrected it to 21.5 kpc, our simulations show that 19.8 kpc provides the best match to the observations.
The orbit varies significantly with distance: in the $\alpha$-$\delta$ space (Right Ascension and Declination), the orbit exhibits an outward-opening trend as the distance increases, and in the $v_{\rm r}$-$\delta$ space, it shows a large deviations near the ends. The observed stars do not lie precisely along a single orbit, but the 19.8 kpc model captures more of them and aligns better with Ibata's orbit \citep{Ibata2024}. This orbit yields an initial Galactocentric position and velocity of the centroid orbit of (8465.9, -14886.2, -16801.9) pc and (-28.9,  94.8, -112.1) $\mathrm {km~s^{-1}}$ , with the final values of (-10687.7, 15023.7, -12655.9) pc and (93.3, -96.3, -102.9) $\mathrm {km~s^{-1}}$, respectively.

\section{The evolution of the virial ratio} \label{sec:A2}
\renewcommand{\thefigure}
{A\arabic{figure}} 
\setcounter{figure}{1}

\begin{figure}[ht!]
    \centering
    \includegraphics[width=1\linewidth]{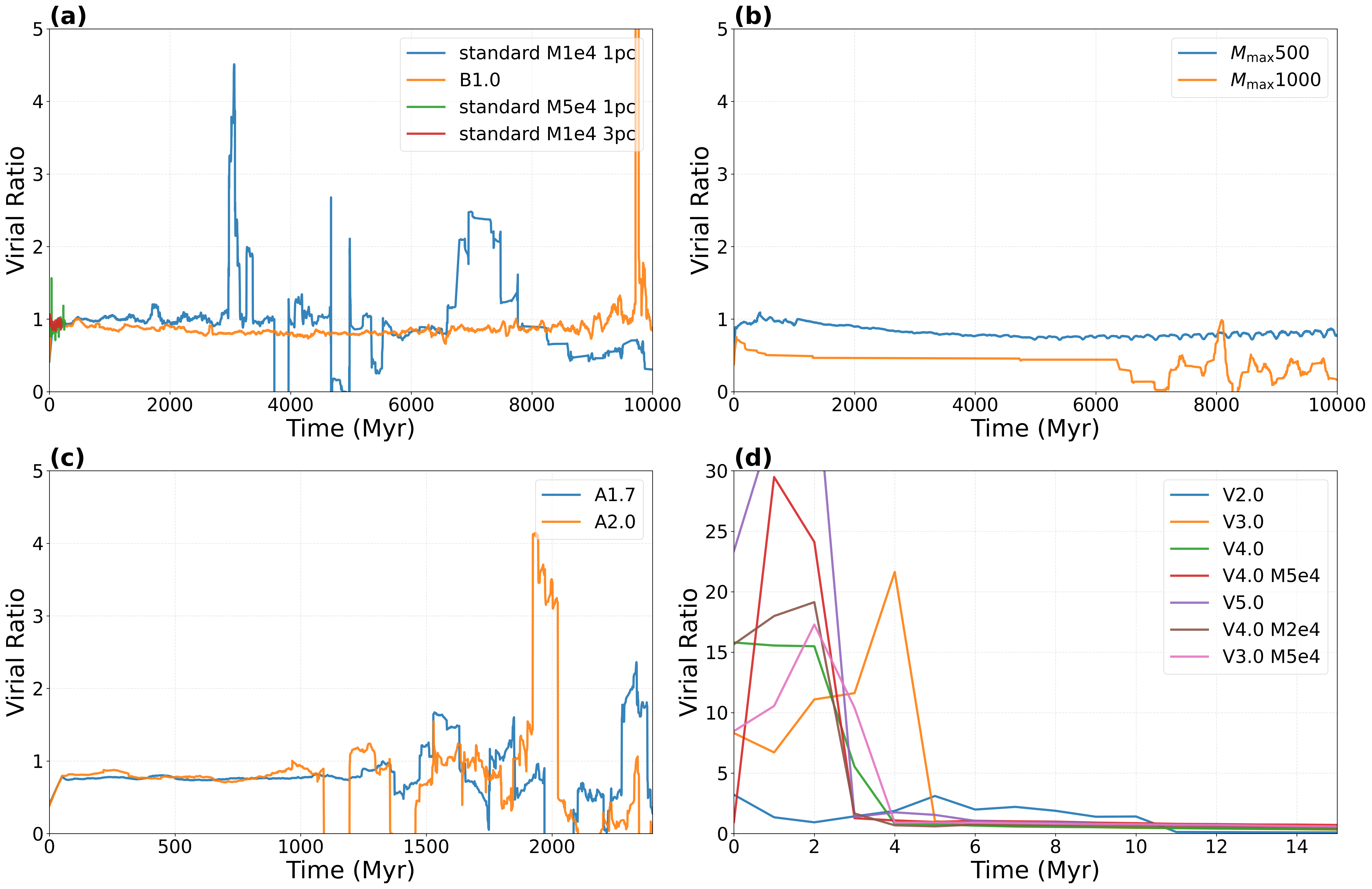}
    \caption{The figure shows the evolution of the virial ratio of star clusters under different models. (a) Virial ratio evolution for standard models with different initial masses and half-mass radii, as well as the binary model, all fluctuating around virial equilibrium. (b) Virial ratio evolution for models with increased maximum stellar mass, which also remain close to virial equilibrium. (c) Virial ratio evolution for power-law models, showing significant fluctuations in early stage, indicating quick dissolution of the system. (d) Virial ratio evolution for gas expulsion models, exhibiting a strongly supervirial state at early times and rapidly declining to low values within a short timescale.
}
    \label{A2}
\end{figure}

We plot the time evolution of the virial ratio for all models described in Section~\ref{sec:models} in Figure~\ref{A2}. Occasional center-of-mass calculation errors, which cause large kinetic energy fluctuations, and limited sampling as clusters approach dissolution produce some extreme, nonphysical peaks. Since the virial ratio is primarily meaningful for bound systems, the time axis is restricted to the evolution prior to complete cluster dissolution. The results show that the standard and binary models maintain virial ratios close to 1 over long timescales. Among the top-heavy IMF models, only the 
$M_{\max}500$ model exhibits a relatively stable virial equilibrium. In contrast, the $M_{\max}1000$ model, due to its higher upper mass limit (with the most massive BH reaching $\sim100~\mathrm{M}\odot$, compared to $\sim10~\mathrm{M}\odot$ in the $M_{\max}500$ model) and enhanced mass segregation, results in a reduced number of stars in the cluster core; as early as $12~\rm{Myr}$, the number of stars within the core radius drops below 10, causing a significant departure of the virial ratio from equilibrium at early times. The two power-law models show strong fluctuations as the systems approach dissolution. The gas expulsion models, on the other hand, exhibit a strongly supervirial initial state and undergo rapid dissolution.

\bibliography{paper}{}
\bibliographystyle{aasjournal}
\end{CJK*}
 \end{document}